\documentclass[final,numberedheadings]{aipproc}
\pdfoutput=1 
\layoutstyle{8x11single}
\usepackage{amsfonts}
\usepackage[centertags]{amsmath}
\usepackage{amssymb,amscd,amsbsy}
\usepackage{eurosym}
\usepackage{./styles/myxspace}
\usepackage{sidecap}
\usepackage[font=small,labelfont=bf,labelsep=period]{caption}
\def\zero{{\scriptscriptstyle 0}}

\catcode`\@=11 
\@ifundefined{euro}%
{\providecommand{\Euro}{Euro\xspace}
 }%
{\providecommand{\Euro}{\euro\xspace}
 }
\catcode`\@=12 

\def\Z0{{Z^\zero}}
\def\eVdist{\kern-0.06667em}

\def\IP{{\rm I$\kern-0.01667em$P}\xspace}

\def\Ptlj{{\not{\kern-0.55ex P}}_t\ell j}
\def\Ptmiss{{\not{\kern-0.55ex P}}_t}

\mathchardef\qsm=63
\mathchardef\pls=43
\mathchardef\mns=512
\mathchardef\plm=518
\mathchardef\eql=61
\mathchardef\smallleft=300
\mathchardef\smallright=301
\mathchardef\perslsh=47
\mathchardef\les=316
\mathchardef\gre=318
\mathchardef\leq=532
\mathchardef\grq=533
\chardef\usc=95
\chardef\til=126
\chardef\amp=38

\def\sqr#1#2#3{{\vcenter{\hrule height.#3ex\hbox{\vrule width.#2ex height#1ex
    \kern#1ex\vrule width.#3ex}\hrule height.#2ex}}}

\def\angleto{\vrule width.035em height2.1ex depth-.56ex\unskip\kern-.6ex\to}
\def\perchc#1{{\raise.4ex\hbox{$\mkern4mu#1{\it\perslsh}_
             {\mkern-5mu\scriptscriptstyle{{\rm o}\!{\rm o}}}^
             {\mkern-12.8mu\scriptscriptstyle{\rm o}}$}}}

\catcode`\@=11 
\def\parenbar{\mathpalette\p@renb@r}
\def\p@renb@r#1#2{\vbox{%
  \ifx#1\scriptscriptstyle \dimen@.7em\dimen@ii.2em\else
  \ifx#1\scriptstyle \dimen@.8em\dimen@ii.25em\else
  \dimen@1em\dimen@ii.4em\fi\fi \offinterlineskip
  \ialign{\hfill##\hfill\cr
    \vbox{\hrule width\dimen@ii}\cr
    \noalign{\vskip-.3ex}%
    \hbox to\dimen@{$\mathchar300\hfil\mathchar301$}\cr
    \noalign{\vskip-.3ex}%
    $#1#2$\cr}}}
\catcode`\@=12 
\def\nuan{\parenbar{\nu}\kern-0.4ex}

\newbox\struttbox
\setbox\struttbox=\hbox{\vrule height1.65ex depth.485ex width0pt}
\def\strutt{\relax\ifmmode\copy\struttbox\else\unhcopy\struttbox\fi}
\def\stru#1#2{\relax\ifmmode\hbox{\vrule height#1 depth#2 width0pt}
\else\vrule height#1 depth#2 width0pt\fi}
\def\uline#1{$\underline{\hbox{#1\strutt}}$}
\def\ronum#1{\uppercase\expandafter{\romannumeral#1}}
\def\ronuml#1{\expandafter{\romannumeral#1}}

\def\cbk{\kern-0.5em}

\newcommand{\linebox}[2][3.ex]{\uline{\hbox to #2{\stru{#1}{0.pt}\hfil}}}

\newcounter{seqnum}
\DeclareMathAlphabet{\mathbf}{OT1}{cmr}{bx}{n}

\DeclareMathAlphabet{\mathbfs}{OT1}{lcmss}{bx}{sl}

\newcommand{\PreserveBackslash}[1]{\let\temp=\\#1\let\\=\temp}
\newlength\listtextwidth

\catcode`\@=11 
\newlength{\@tabfninsert}
\newlength{\@tabfnwidth}
\newcommand{\tabfootnote}[2]{%
  \setlength{\@tabfninsert}{0.8em}
  \setlength{\@tabfnwidth}{\textwidth}
  \addtolength{\@tabfnwidth}{-\@tabfninsert}
  \addtolength{\@tabfnwidth}{-0.4em}
  \noindent\makebox[\@tabfninsert][r]{\footnotesize$^{#1}$\hfil}\hfill%
  \parbox[t]{\@tabfnwidth}{\footnotesize #2\hfill}}
\catcode`\@=12 
\newcommand{\boldarrayrulewidth}{1pt}

\catcode`\@=11 
\let\tab@penalty\relax
\newcount\tab@state
\def\bcline#1{%
  \noalign{\kern-.5\arrayrulewidth\tab@penalty}%
  \omit%
  \global\tab@state\@ne%
  \ranges\bcline@i{#1}%
  \cr%
  \noalign{\kern-.5\arrayrulewidth\tab@penalty}%
}
\def\bcline@i#1#2{%
  \ifnum#1<\tab@state\relax%
    \tab@@cr%
    \noalign{\kern-\arrayrulewidth\tab@penalty}%
    \omit%
    \global\tab@state\@ne%
  \fi%
  \@whilenum\tab@state<#1\do{%
    \hfil\tab@@tab@omit%
    \global\advance\tab@state\@ne%
  }%
  \ifnum\tab@state>\@ne%
    \kern-\arrayrulewidth%
  \fi%
  \@whilenum\tab@state<#2\do{%
    \tab@@span@omit%
    \global\advance\tab@state\@ne%
  }%
  \leaders\hrule\@height\boldarrayrulewidth\hfill%
}
\def\ranges#1#2{%
  \gdef\ranges@temp{#1}%
  \begingroup%
  \ranges@i#2 \q@delim%
}
\def\ranges@i{%
  \@ifnextchar\q@delim\ranges@done{\afterassignment\ranges@ii\count@}%
}
\def\ranges@ii{%
  \@ifnextchar-\ranges@iii{\ranges@do\count@\count@\ranges@v}%
}
\def\ranges@iii-{\afterassignment\ranges@iv\@tempcnta}
\def\ranges@iv{\ranges@do\count@\@tempcnta\ranges@v}
\def\ranges@v{%
  \@ifnextchar,%
    \ranges@vi%
    {%
      \@ifnextchar\q@delim%
        \ranges@done%
        {\tab@err@range\ranges@vi,}%
    }%
}
\def\ranges@vi,{\afterassignment\ranges@ii\count@}
\def\ranges@do#1#2{%
  \ifnum#1>#2\else%
    \expandafter\endgroup%
    \expandafter\ranges@temp%
    \expandafter{%
    \the\expandafter#1%
    \expandafter}%
    \expandafter{%
    \the#2%
    }%
    \begingroup%
  \fi%
}
\def\ranges@done\q@delim{\endgroup}
\def\ifinrange#1#2{%
  \@tempswafalse%
  \count@#1%
  \ranges\ifinrange@i{#2}%
  \if@tempswa%
    \expandafter\@firstoftwo%
  \else%
    \expandafter\@secondoftwo%
  \fi%
}
\def\ifinrange@i#1#2{%
  \ifnum\count@<#1 \else\ifnum\count@>#2 \else\@tempswatrue\fi\fi%
}
\def\tab@@cr{\cr}
\def\tab@@tab@omit{&\omit}
\def\tab@@span@omit{\span\omit}
\def\tab@checkrule#1{%
  \count@#1\relax%
  \expandafter\ifinrange%
  \expandafter\count@%
  \expandafter{\tab@xcols}%
    {\tab@checkrule@i}%
    {}%
}
\def\bhline{\noalign{\ifnum0=`}\fi\hrule \@height  
\boldarrayrulewidth \futurelet \@tempa\@xhline}
\def\@xhline{\ifx\@tempa\hline\vskip \doublerulesep\fi
      \ifnum0=`{\fi}}
\catcode`\@=12 

\catcode`\@=11 
\newcounter{pict@width}
\newcounter{pict@height}
\newlength{\pict@scale}
\newcommand{\psfigadd}[4]{%
\setcounter{pict@width}{1*\ratio{#2+\pict@scale/2}{\pict@scale}}
\setcounter{pict@height}{1*\ratio{#3+\pict@scale/2}{\pict@scale}}
\setlength{\unitlength}{\pict@scale}
\hbox to #2{\hspace{-\fill}\begin{picture}(\thepict@width,\thepict@height)
\put(0,0){\psfig{figure=#1,width=#2,height=#3,clip=}}
\SetScale{0.283466457}
\SetWidth{1.763889}
{#4}
\end{picture}}
}
\newcounter{pict@widthfst}
\newcounter{pict@widthscd}
\newcounter{pict@widthtot}
\newcommand{\psfigaddtwo}[7]{%
\setcounter{pict@widthfst}{1*\ratio{#2+\pict@scale/2}{\pict@scale}}
\setcounter{pict@widthscd}{1*\ratio{#2+#4+\pict@scale/2}{\pict@scale}}
\setcounter{pict@widthtot}{1*\ratio{#2+#4+#6+\pict@scale/2}{\pict@scale}}
\setcounter{pict@height}{1*\ratio{#3+\pict@scale/2}{\pict@scale}}
\setlength{\unitlength}{\pict@scale}
\hbox{\hspace{-\fill}\begin{picture}(\thepict@widthtot,\thepict@height)
\put(0,0){\psfig{figure=#1,width=#2,height=#3,clip=}}
\put(\thepict@widthscd,0){\psfig{figure=#5,width=#6,height=#3,clip=}}
\SetScale{0.283466457}
\SetWidth{1.763889}
{#7}
\end{picture}}
}
\newcommand{\psfigror}[4]{%
\setcounter{pict@width}{1*\ratio{#2+\pict@scale/2}{\pict@scale}}
\setcounter{pict@height}{1*\ratio{#3+\pict@scale/2}{\pict@scale}}
\setlength{\unitlength}{\pict@scale}
\hbox{\begin{picture}(\thepict@width,\thepict@height)
\put(0,\thepict@height){\psfig{figure=#1,width=#3,height=#2,clip=,angle=270}}
\SetScale{0.283466457}
\SetWidth{1.763889}
{#4}
\end{picture}}
}
\newcommand{\psfigrol}[4]{%
\setcounter{pict@width}{1*\ratio{#2+\pict@scale/2}{\pict@scale}}
\setcounter{pict@height}{1*\ratio{#3+\pict@scale/2}{\pict@scale}}
\setlength{\unitlength}{\pict@scale}
\hbox{\begin{picture}(\thepict@width,\thepict@height)
\put(0,0){\psfig{figure=#1,width=#3,height=#2,clip=,angle=90}}
\SetScale{0.283466457}
\SetWidth{1.763889}
{#4}
\end{picture}}
}
\catcode`\@=12 

\begin{document}
\title{News from KM3NeT}
\author{Ulrich F.\,Katz for the KM3NeT Collaboration}{address=
       {Friedrich-Alexander University of Erlangen-Nürnberg,\\ 
        Erlangen Centre for Astroparticle Physics\\
        E-mail: katz@physik.uni-erlangen.de}}

\begin{abstract}
KM3NeT is a future research infrastructure in the Mediterranean Sea, hosting a
multi-cubic-kilometre neutrino telescope and nodes for Earth and Sea sciences. 
In this report we shortly summarise the genesis of the KM3NeT project and
present key elements of its technical design. The physics objectives of the
KM3NeT neutrino telescope and some selected sensitivity estimates are discussed. 
Finally, some first results from prototype operations and the next steps towards
implementation -- in particular the first construction phase in 2014/15 -- are
described.
\end{abstract}
\maketitle
\section{Introduction: KM3NeT Genesis}
\label{sec:int}

Neutrino telescopes to measure high-energy cosmic neutrinos as messengers for
astrophysical observations have been suggested more than half a decade ago (see
\cite{Katz+Spiering-2012} for a detailed review of neutrino telescopes and for
further references). In the mid-1990s the first instruments were in operation
(Baikal, AMANDA) and demonstrated the viability of the detection principle. At
the same time, it became clear from their first results and refined theoretical
studies that cubic-kilometre installations will be needed to exploit the physics
potential of neutrino astronomy. This led to the construction of the IceCube
telescope at the South Pole, while an instrument of similar size was recommended
for the Northern hemisphere to complement IceCube in its field of view
\cite{HENAP-2002}. In 2006--2009, the Mediterranean neutrino telescope
collaborations (ANTARES, NEMO, NESTOR) have conducted a 3-year design
study\footnote{Co-funded through the 6th Framework Programme of the EU, contract
no.~011937} to work out the design of such an instrument complying with the
sensitivity and reliability requirements and within a cost frame of 200\,M\Euro
per cubic kilometre of instrumented water. In 2010, the resulting technical
design was presented in \cite{KM3NeT-TDR-2010}; at that time, various options of
technical solutions were described and final decisions on technology and site(s)
were still pending; a high level of cost effectiveness has been reached, with an
investment volume of 220--250\,M\Euro for 4--5 cubic kilometres of instrumented
volume.

Meanwhile, after an EU-funded Preparatory Phase\footnote{2008--2012, FP7 Grant
Agreement~212525}, the KM3NeT Collaboration has been formally founded with a
governing board, project management, an external scientific and technological
advisory committee and regulations laid down in a Memorandum of Understanding
that is awaiting signature. All pending decisions have been taken (see
Section~\ref{sec:tec}), prototype tests are in progress (Section~\ref{sec:pro})
and a first phase of construction has started. (Sect.~\ref{sec:nex}). Some
examples of expected physics sensitivities are discussed in Sect.~\ref{sec:phy}.

\section{The KM3NeT Technical Design}
\label{sec:tec}

\begin{figure}[thb]
  \includegraphics[width=0.68\textwidth]{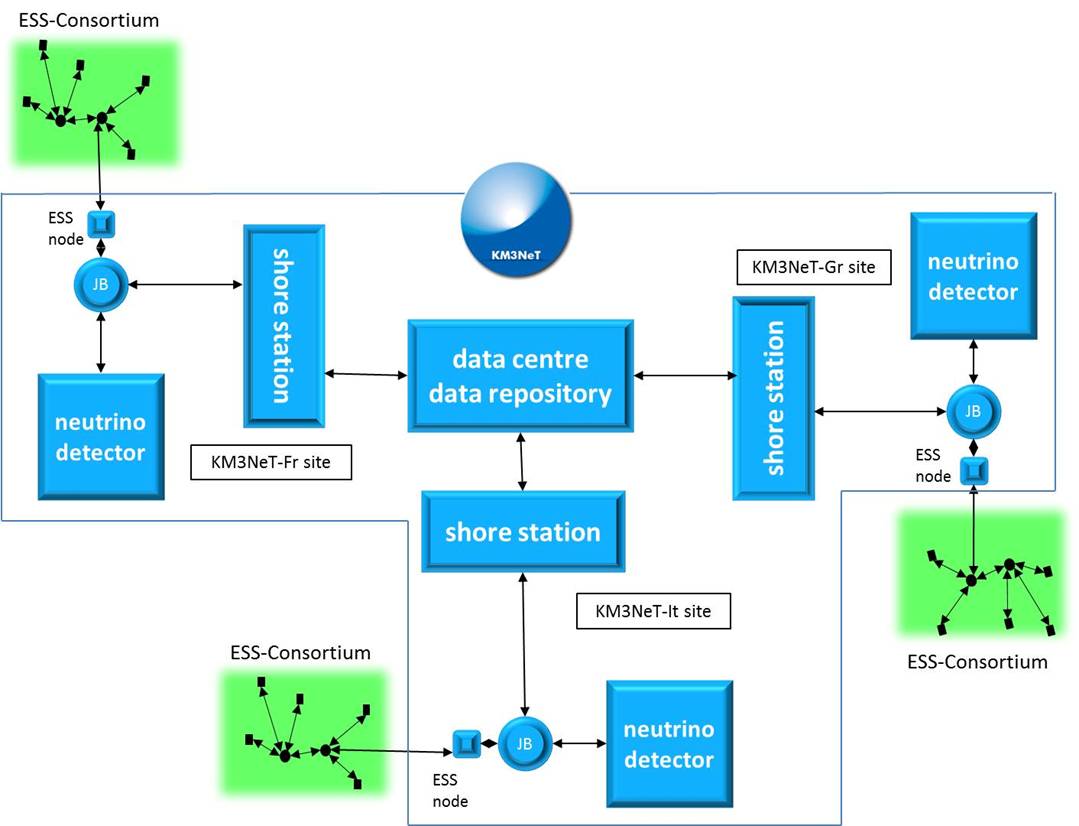}
  \caption{Schematic installation scheme for the KM3NeT research infrastructure.
           The KM3NeT Collaboration plans to provide nodes for the Earth and Sea
           Science (ESS) instrumentation but will not construct the latter.}
  \label{fig:sites}
\end{figure}

The KM3NeT research infrastructure will be constructed as a distributed
installation with common detector technology, management, data handling and
operation control. The installation sites will be close to Toulon, France
(KM3NeT-Fr), near the East coast of Sicily, Italy (KM3NeT-It), and West of the
Peloponnesus, Greece (KM3NeT-Gr) (see Fig.~\ref{fig:sites}). This decision is
based on detailed simulation studies indicating that -- at least for
muon-neutrino signals from Galactic point sources -- the overall sensitivity is
not reduced by splitting the detector in independent building blocks, provided
these have a minimal size of about $0.5\,\text{km}^3$. The rationale for several
building blocks, as opposed to a large homogeneous detector, is both technical
(related to the deep-sea cable network, power and data bandwidth requirements,
installation/deployment issues and reliability of the detector during operation)
and funding-related (to secure a significant amount of regional funding). Also,
the distributed installation addresses the interest of the Earth and Sea
sciences to establish a multitude of observation sites

\begin{figure}[hbt]
  \includegraphics[height=0.28\textheight]{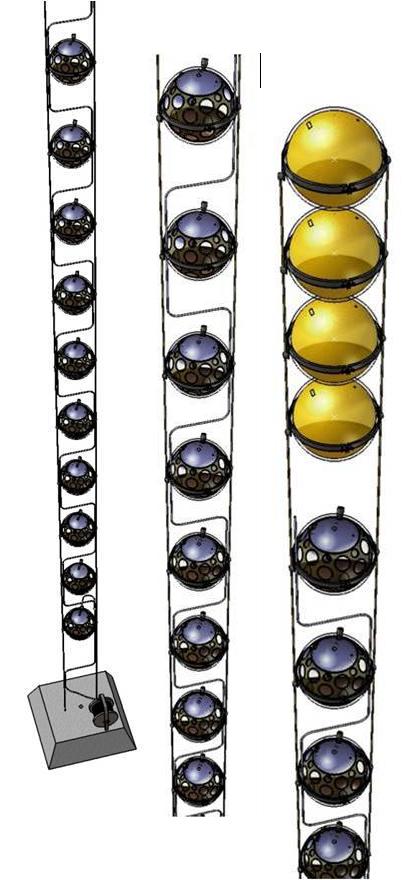}\kern1.5cm
  \includegraphics[height=0.28\textheight]{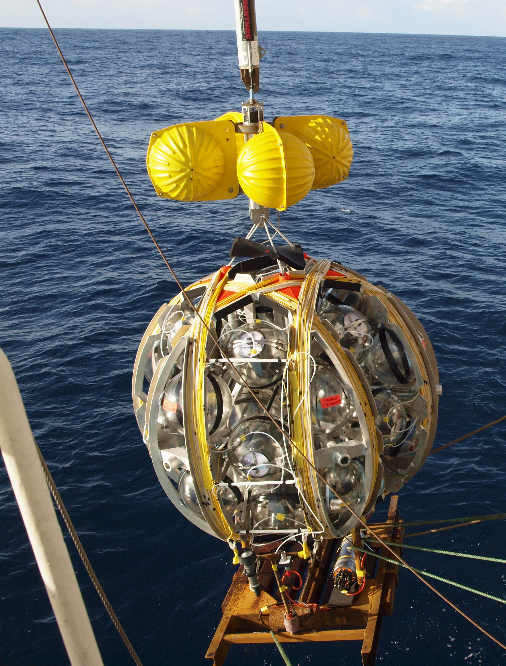}\kern1.5cm
  \includegraphics[height=0.28\textheight]{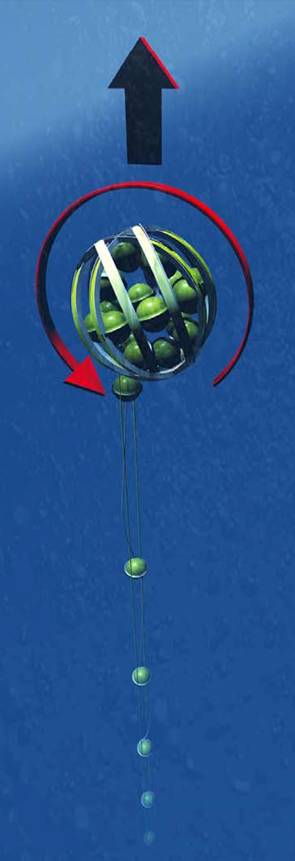}
  \caption{Left: Artist's view of a DU, split into 3 parts (anchor/middle/top) 
           for presentational simplicity. The VEOC is attached in meanders along 
           the DU to avoid tensile load and asymmetric drag in sea currents. Note 
           that the picture is not to scale. Middle: A loaded LOM during a deployment
           test. Right: Upward rotational motion of the LOM during DU unfurling
           (picture by courtesy of Marijn van de Meer, Quest).}
  \label{fig:du}
\end{figure}

At each site, two building blocks of 115~{\it Detection Units (DUs)} (see
Fig.~\ref{fig:du} left) each will be constructed. The average distance between
neighbouring DUs is 90\,m. Each DU carries 18~{\it Digital Optical Modules
(DOMs)}, starting 100\,m above the sea floor and with 36\,m distance between
adjacent DOMs. The DUs are supported by two pre-stretched Dyneema$^\copyright$
ropes and kept straight by a submerged buoy at their top. The vertical
electro-optical cable (VEOC) consists of a flexible, oil-filled hose that is in
equi-pressure with the sea water and contains optical fibres for data transport
and copper wires for electrical power provision. For each DOM, a break-out box
provides connection to one fibre and two wires. Overall, this design minimises
the number of pressure transitions and allows for a slim, light-weight
construction with moderate drag in sea currents. At a current of 30\,cm/s,
corresponding to the maximum observed at any of the installation sites, the DU
top is deflected from its nominal position by about 150\,m, which is well inside
tolerances (note that the currents are expected to be homogeneous over the
volume of one detector block).

For deployment, a DU will be wrapped on a spherical frame with about 2.2\,m
diameter ({\it Launching vehicle for Optical Modules, LOM}) \cite{lom-2013}
which is deposited on the seabed and then unfurls in a rotating upwards movement
(see Fig.~\ref{fig:du} middle and right). The LOM rises to the sea surface,
where it is collected for reuse.

\begin{figure}[hbt]
  \includegraphics[height=0.23\textheight]{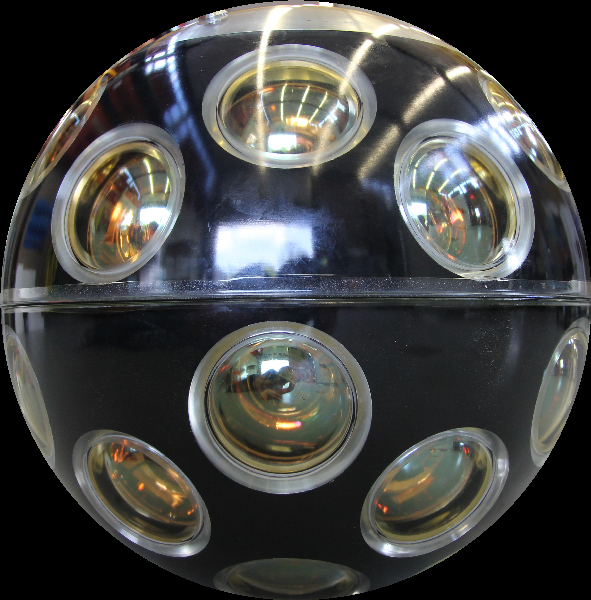}\kern1.5cm
  \includegraphics[height=0.23\textheight]{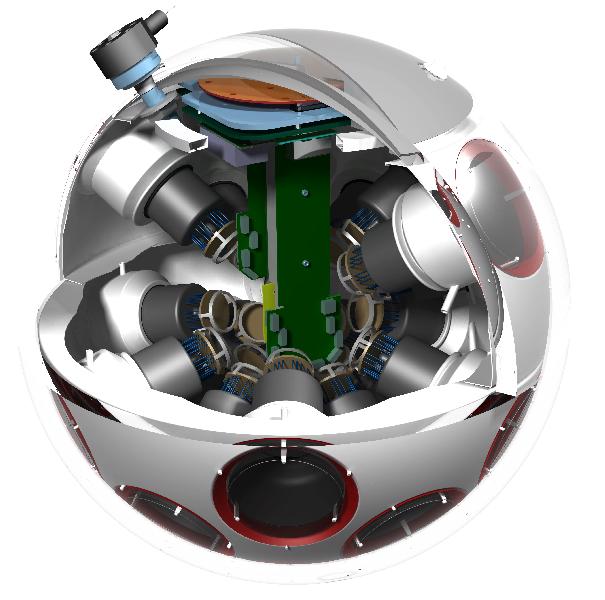}
  \caption{Photograph of a prototype (left) and technical drawing (right) of a 
           KM3NeT DOM. Note that the DOM contains all front-end electronics
           components and also a aluminum structure for conducting the heat to
           the glass surface.}
  \label{fig:dom}
\end{figure}

Each DOM (see Fig.~\ref{fig:dom}) is a pressure-resistant glass sphere of 17
inch diameter that carries 31 3-inch photomultiplier tubes (PMTs) with their
high-voltage bases as well as calibration devices and readout electronics. The
lower hemisphere of each DOM contains 19 of the PMTs, which are thus
``downward-looking'', whereas the other 12 PMTs look upwards. PMTs from two
companies (Hamamatsu, ETEL) have been demonstrated to fulfil the KM3NeT
specifications \cite{vlvnt13-kalekin,vlvnt13-leonora} and can be produced in the
required quantities in due time. The use of reflective rings around the PMT
faces increases the light collection efficiency per PMT by about 27\%
\cite{km3net-econe-2013}. All PMTs are controlled and read out individually. The
DOM contains front-end electronics to amplify the PMT signals and to transform
them into digital time-over-threshold information that is fed into the readout
via optical fibres (see \cite{vlvnt13-real} for details on electronics and data
acquisition). All PMT signals above an adjustable noise threshold (typically the
equivalent of 0.3\;photo-electrons) are sent to shore, where event candidates
are selected by online filters running on a computer farm. The main advantages
of the multi-PMT design are: (i) photocathode area per DOM increased by a factor
of three as compared to optical modules with one 10-inch photomultiplier; (ii)
segmented photocathode, allowing for unambiguous recognition of coincident hits
and providing some directional sensitivity; (iii) reduced cost and risk due to
fewer optical modules, fewer pressure transitions, and simplified mechanical
structure; (iv) almost $4\pi$ solid angle coverage by each DOM.

The time synchronisation of the DOMs at nanosecond precision is achieved by a
combination of the White Rabbit \cite{whiterabbit-2013} ethernet protocol for
the shore-to-DOM communication and laser \cite{vlvnt13-real2} and LED
\cite{vlvnt13-calvo} flashers in the neutrino telescope. The position
calibration of each DOM is achieved at about 10\,cm precision using acoustic
triangulation. The acoustic system includes transponders at the seafloor and a
receiver in each DOM. These sensors use piezo elements that are integrated with
the digitisation electronics and can thus be operated in electromagnetically
noisy environments \cite{vlvnt13-enzen}. The DOM orientations are monitored at a
few degrees using compass and tilt sensors in the DOMs.

\section{Physics Perspectives}
\label{sec:phy}

As recently impressively demonstrated by the first detection of high-energy
cosmic neutrinos with the IceCube detector \cite{icecube-2013}, neutrino
telescopes are sensitive to neutrinos from all directions, albeit not equally
for all reaction channels and energies. Detecting the muons produced in $\nu_\mu$
charged current reactions\footnote{Here and in the following, $\nu$ is
understood to denote both neutrinos and antineutrinos.}, $\nu_\mu N\to\mu X$
($N$ being the target nucleon and $X$ the hadronic final state), provides the
best directional resolution and therefore the best prospects to identify cosmic
neutrino sources. At neutrino energies up to some $100\,$TeV, the sensitivity
for this channel is best for upward-going neutrinos; due to its location in the
Mediterranean Sea, KM3NeT is therefore in an optimal position to investigate
neutrino fluxes from the Southern sky, in particular from the Galactic Centre
and the largest fraction of the Galactic plane.

\begin{SCfigure}[1.0][thb]
  \includegraphics[width=0.6\textwidth]{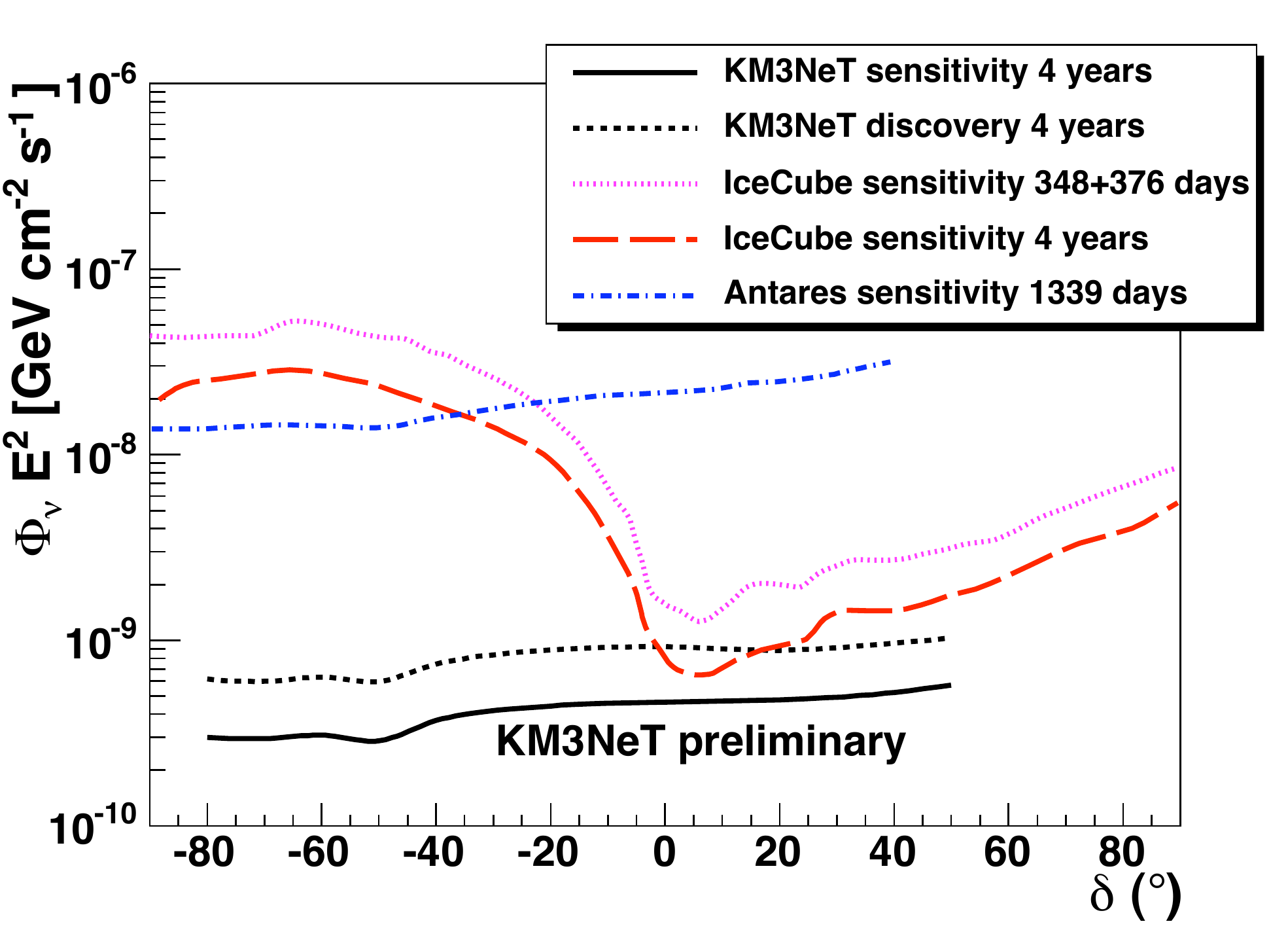}
  \caption{\ \ KM3NeT $5\sigma$ discovery potential and sensitivity for point
           sources emitting a muon neutrino flux with an $E^{-2}$ spectrum, for
           4 years of data taking with the full detector configuration. The flux
           values are shown as function of the source declination. For
           comparison the corresponding sensitivities are also shown for ANTARES
           and IceCube (figure from \cite{vlvnt13-trovato}).}
\label{fig:points}
\end{SCfigure}

Identifying Galactic neutrino sources is a priority physics goal with the full
KM3NeT detector of six building blocks (see Sect.~\ref{sec:tec}). Simulation
studies \cite{vlvnt13-trovato}\footnote{Note that these studies have been done
for a previously considered detector configuration with two large building
blocks with 310 DUs each and inter-DU distances of 100\,m; the differences with
respect to the 6-block configuration are, however, expected to be small.} show
that this detector will provide a sensitivity exceeding and complementing that
of IceCube for a large range in declination (see Fig.~\ref{fig:points}). 
Specific Galactic sources with a cut-off in their energy spectrum, as derived
from high-energy gamma-ray measurements, have also been considered; it was e.g.\
found that the Supernova remnant RX\,J1713.7-3946 can be detected with $5\sigma$
in about 5~years if its gamma-ray emission is of purely hadronic origin.

As an example of an extended emission region, the KM3NeT sensitivity to a
possible neutrino flux from the ``Fermi Bubbles'' (two large regions above and
below the Galactic Centre with a hard-spectrum gamma-ray emission detected by
the Fermi satellite) has been investigated
\cite{km3net-fermibubbles-2013,vlvnt13-coniglione}. Again, neutrino fluxes from
these regions are readily detectable with KM3NeT if the underlying processes are
hadronic and the spectrum extends up to at least $30\,$TeV.

Due to the focus on (Galactic) point sources, the detection of other than
$\nu_\mu$ charged current reaction channels with KM3NeT is still under study. In
particular, the sensitivity to a cosmic neutrino signal as measured by IceCube
\cite{icecube-2013} is expected to be reported soon.

A case study is pursued by the KM3NeT collaboration on the option of employing
KM3NeT technology for a dense, low-energy detector; the main objective would be
a precision measurement of atmospheric neutrinos in the energy range of about
$5\text{--}20\,$GeV and a determination of the neutrino mass hierarchy from the
observed distribution of events in energy and declination (ORCA project
\cite{vlvnt13-tsirigotis,neutel13-orca}).

\section{First results from Prototype Tests}
\label{sec:pro}

Using a prototype LOM and a full-length DU with spheres, ropes and the vertical
cable (VEOC), but without photo-detection or calibration instrumentation,
deployment tests have been performed in April 2013. A problem with a penetrator
was found and solved by redesign; various smaller issues have been identified
that led to modifications of a few details of DU mechanics and LOM. Altogether
the tests have clearly demonstrated that the procedure is feasible and safe, and
that the DUs deployed behaved as expected (no twisting, appropriate orientation
of DOMs). A further set of deployment tests with modified LOM and DU design is
planned for early summer 2014.

\begin{figure}[hbt]
  \includegraphics[width=0.49\textwidth]{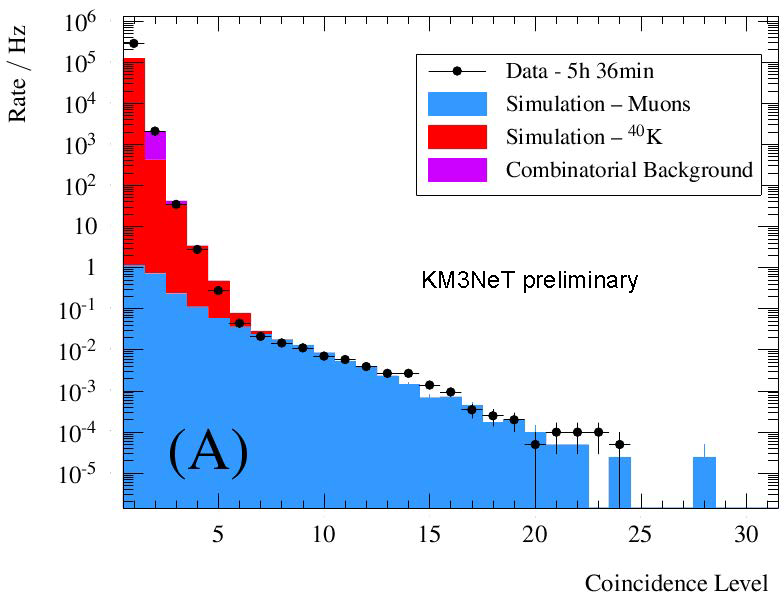}\hfill
  \includegraphics[width=0.49\textwidth]{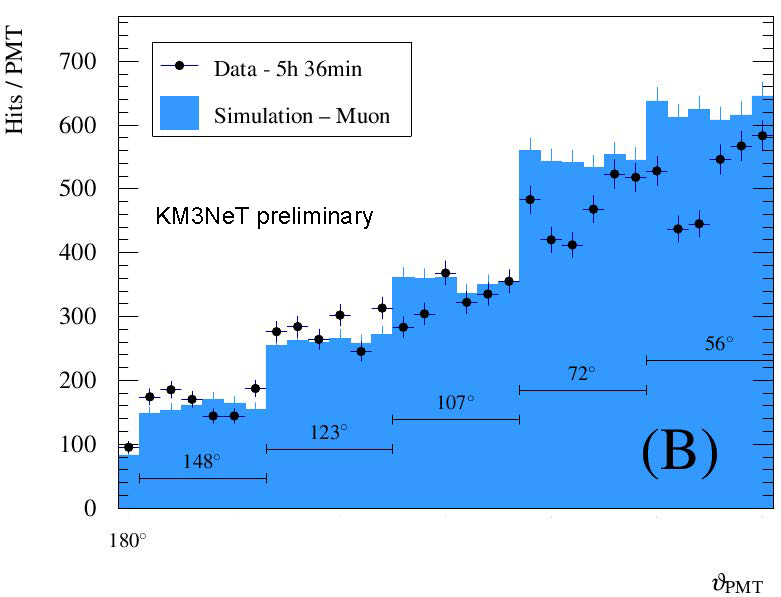}
  \caption{(A) Distribution of the coincidence levels in the prototype DOM,
           compared to simulation results including light from potassium-40
           decays, and atmospheric muons, plus combinatorial background. Note
           that events with signals on more than 20 out of 31 PMTs are observed. 
           (B) Zenith angle $\theta_\text{PMT}$ of the PMT axis for those PMTs
           that are hit in events with more than 7 coincidences. The
           upward-looking PMTs observe clearly larger rates than the
           downward-looking ones, consistent with the angular spectrum of the
           muons. The discrepancies for a few PMTs at lower values of
           $\theta_\text{PMT}$ are attributed to shadowing by mechanical
           structures (not yet included in the simulation). Figures taken from
           \cite{vlvnt13-michael}.}
  \label{fig:ppmdom}
\end{figure}

A prototype of a KM3NeT DOM, fully equipped with photomultipliers and adapted to
the ANTARES data acquisition, has been attached to an upgraded instrumentation
line of ANTARES and deployed early 2013. Data taking is proceeding since April
2013. The DOM is found to be operating fully within specifications
\cite{vlvnt13-michael}. The results
are in very good agreement with simulations and demonstrate the versatility of
the multi-PMT concept as well as the richness and quality of information
contained in the data. As an example, Fig.~\ref{fig:ppmdom} shows the
distribution of the coincidence levels (i.e.\ the number of PMTs with coincident
signals) and the zenith angle distribution of hit PMTs in events with more than
seven coincidences. It is obvious that one single DOM is able to identify muon
events and to address their angular distribution, an achievement that is
unthinkable for optical modules with one large PMT.

\section{Next Steps}
\label{sec:nex}

The KM3NeT collaboration currently pursues the completion of the qualification
tests and, in parallel, the preparation of a first construction phase at the
KM3NeT-Fr and KM3NeT-It sites. As a next qualification step after the prototype
DOM, a reduced-size DU with three DOMs and the final electronics will be
deployed at the KM3NeT-It site to test, amongst others, inter-DOM
synchronisation, readout, data acquisition and the connection to the deep-sea
cable network. The deployment is expected for the first half of 2014.

According to the current planning, the production, deployment and connection of
31~DUs will take place in the period 2014 to 2016. This time scale is enforced
by constraints on the availability of regional funding resources that need to be
spent in accordance with EU rules. All detector components are scrutinised in
{\it production readiness reviews} conducted by external panels in the first
half of 2014. The first regular DU will be deployed at KM3NeT-Fr and will serve
as a concluding engineering test. Of the remaining DUs, 24 will be deployed at
KM3NeT-It to form the most sensitive neutrino telescope in the Northern
hemisphere, with roughly 15--20\% of the IceCube sensitivity\footnote{A
prototype tower following the Italian ``flexible tower'' concept
\cite{KM3NeT-TDR-2010} has been deployed and is in operation, providing data
that confirm the low bioluminescence level observed in previous measurement
campaigns at the KM3NeT-It site.}.

The next, intermediate objective is the construction of two full building
blocks, providing a sensitivity similar to that of IceCube. This installation,
which could be completed around 2017/18 if funding is secured, will have as a
central physics focus the investigation of the characteristics and sources of
the cosmic neutrino flux detected by IceCube.

Full KM3NeT construction (six building blocks, altogether 3--4\,km$^3$) will
proceed subsequently.

Discussions are ongoing on how to include ORCA into this scenario.

\section{Conclusions and Outlook}
\label{sec:con}

After about a decade of technical R\&D work and simulation studies addressing
the physics capabilities, the KM3NeT collaboration has converged on the
technical design for the future, multi-km$^3$ KM3NeT neutrino telescope in the
Mediterranean Sea. The main design concepts are the digital multi-PMT optical
module (DOM); slim, flexible detection units with an equi-pressure backbone
cable; their deployment on spherical structures that unfurl the DUs while
``rolling up'' from the sea bed. All photomultiplier tubes are read out
individually; the time-over-threshold information of each PMT pulse is digitised
in the DOM. Optical point-to-point connections between each DOM and shore are
used for data transport and time synchronisation, allowing for sending all PMT
data to shore, where they are filtered for event candidates by an online
computer farm. The detector components are currently under final scrutiny,
and first prototype tests in the deep sea yield very encouraging results.

KM3NeT will be constructed as a distributed research infrastructure at sites
in France, Italy and Greece. Eventually, two building blocks with 115 DUs each
are foreseen for each site. Operation, computing and governance are organised
centrally, and the technical design of the DUs including aspects such as data
acquisition and data formats is identical at these sites. At all sites, KM3NeT
will provide access to real-time, continuous measurements of Earth and Sea
sciences.

Currently a first construction phase at the KM3NeT-Fr and KM3NeT-It sites is
funded and in preparation. A staged approach is pursued towards the completion
of KM3NeT, with two building blocks as an intermediate milestone to reach an
IceCube-like sensitivity.

\section*{Acknowledgements}

The author would like to thank the organisers for a truly inspiring workshop in
a wonderful setting. The KM3NeT project has been strongly supported by the EU
through the FP6 Design Study (contract no.\ 011937) and the FP7 Preparatory
Phase (grant agreement no.\ 212525).

{
\bibliographystyle{./VLVnT}
{\raggedright\fontsize{10.pt}{11.pt}\selectfont
\bibliography{./VLVnT}}
}

\end{document}